\documentclass[sigconf,screen,natbib=false]{acmart}

\AtBeginDocument{%
  \providecommand\BibTeX{{%
    \normalfont B\kern-0.5em{\scshape i\kern-0.25em b}\kern-0.8em\TeX}}}

\setcopyright{acmlicensed}
\copyrightyear{2024}
\acmYear{2024}
\acmDOI{XXXXXXX.XXXXXXX}

\acmConference[FSE'24]{ACM Symposium on the Foundations of Software Engineering}{November 15--19, 2024}{Porto de Galinhas, Brazil}
\acmBooktitle{Proceedings of the 32nd ACM Symposium on the Foundations of Software Engineering (FSE '24), November 15--19, 2024, Porto de Galinhas, Brazil}
\acmISBN{978-1-4503-XXXX-X/18/06}

\RequirePackage[
  datamodel=acmdatamodel,
  style=acmnumeric,
  natbib,
  backend=biber
  ]{biblatex}
\bibliography{bibliography}
\usepackage{booktabs}
\usepackage{hyperref}
\usepackage{url}
\usepackage{todonotes}
\usepackage{placeins}
\begin{document}

\title{QCSHQD: Quantum computing as a service for Hybrid classical-quantum software development: A Vision} 

\author{Maryam Tavassoli Sabzevari}
\email{maryam.tavassolisabzevari@oulu.fi}
\orcid{0009-0004-8879-0285}
\affiliation{
  \institution{University of Oulu}
  \streetaddress{Pentti Kaiteran Katu, 1}
  \city{Oulu}
  \country{Finland}
  \postcode{FI-90550}
}

\author{Matteo Esposito}
\email{m.esposito@ing.uniroma2.it}
\orcid{0000-0002-8451-3668}
\affiliation{
  \institution{University of Rome  ``Tor Vergata''}
  \streetaddress{Via del Politecnico, 1}
  \city{Rome}
  \state{Lazio}
  \country{Italy}
  \postcode{00132}
}

\author{Davide Taibi}
\email{davide.taibi@oulu.fi}
\orcid{0000-0002-3210-3990}
\affiliation{
  \institution{University of Oulu}
  \streetaddress{Pentti Kaiteran Katu, 1}
  \city{Oulu}
  \country{Finland}
  \postcode{FI-90550}
}

\author{Arif Ali Khan}
\email{arif.khan@oulu.fi}
\orcid{0000-0002-8479-1481}
\affiliation{
  \institution{University of Oulu}
  \streetaddress{Pentti Kaiteran Katu, 1}
  \city{Oulu}
  \country{Finland}
  \postcode{FI-90550}
}


\begin{abstract}
Quantum Computing (\textbf{QC}) is transitioning from theoretical frameworks to an indispensable powerhouse of computational capability, resulting in extensive adoption across both industrial and academic domains. QC presents exceptional advantages, including unparalleled processing speed and the potential to solve complex problems beyond the capabilities of classical computers. Nevertheless, academic researchers and industry practitioners encounter various challenges in harnessing the benefits of this technology. The limited accessibility of QC resources for classical developers, and a general lack of domain knowledge and expertise, represent insurmountable barrier, hence to address these challenges, we introduce a framework- Quantum Computing as a Service for Hybrid Classical-Quantum Software Development (QCSHQD), which leverages service-oriented strategies. Our framework comprises three principal components: an Integrated Development Environment (IDE) for user interaction, an abstraction layer dedicated to orchestrating quantum services, and a service provider responsible for executing services on quantum computer. 
This study presents a blueprint for QCSHQD, designed to democratize access to QC resources for classical developers who want to seamless harness QC power. The vision of QCSHQD paves the way for groundbreaking innovations by addressing key challenges of hybridization between classical and quantum computers.
\end{abstract}

\ccsdesc[500]{Theory of computation~Design and analysis of algorithms}
\ccsdesc[500]{Theory of computation~Tree languages}

\ccsdesc[500]{Hardware~Quantum technologies}
\ccsdesc[500]{Hardware~Quantum computation}
\ccsdesc[500]{Hardware~Emerging technologies}

\keywords{Quantum computing; Quantum services; Hybrid Classical-quantum; Hybrid Framework}


\maketitle

\section{Introduction}
\label{sec:intro}
Quantum computing has received rapidly growing industrial interest. It is evidenced by the fact that technology giants such as IBM \cite{IBM}, Google \cite{Google}, and Microsoft \cite{Microsoft} have heavily invested in and committed to QC. Moreover, the European Union has invested an expected budget of one billion euros over ten years in an initiative known as the European Quantum Technologies Flagship, to position Europe as a leading force in the quantum revolution\footnote{\url{https://tinyurl.com/mr4bxmv5}}. 

Tech giant and government interest has empowered QC and paved the way for business development in various areas, such as cybersecurity, financial services, healthcare and aerospace \cite{aithal2023advances}.  However, the development of QC applications in a systematic, anticipated, and cost-effective way is critically challenging and remains an unsolved problem
\cite{ali2022software,khan2023software,weder2022quantum,zhao2020quantum}. The need for advanced approaches to tackle this problem is essential to enable QC to live up to its promising potential, therefore, we need advanced solutions to enable QC reaching its full potential. 

Like the classical computing systems, there is a need for quantum software systems and applications that can exploit the benefits of quantum information processing by operationalizing quantum computers \cite{khan2023software}. This need is explicitly highlighted in a national agenda for software engineering research and development presented by Software Engineering Institute (SEI)-Carnegie Mellon University \cite{carleton2021architecting}:\textit{"the development of quantum computing software systems is considered a pivotal future focus area. This paper presents the objectives of this research area, which are initially focused on facilitating the development of quantum software for existing quantum computers and then increasing abstraction as larger, fully fault-tolerant quantum computing systems become available. A key challenge is to fully integrate these types of systems into a unified classical and quantum software development life cycle".} Ensuring robust support for software components is crucial for harnessing the comprehensive capabilities of QC. Therefore, the seamless integration of quantum hardware and software emerges as an essential requirement for fully leveraging the potential of QC technologies \footnote{\url{https://tinyurl.com/2p9y2a64}}. In this study, we presented a Service for Hybrid Classical-Quantum Software Development (QCSHQD) to bridge the gap between classical developers and QC resources by exploiting service-oriented computing strategies.

The remaining sections of this paper are organized as follows: Section \ref{sec:motivation} provides the foundational background and underlying motivations of this idea. Our vision and its workflow are described in Section \ref{sec:vision}. Section \ref{sec:Roadmap} discusses the tools and technologies selected for implementing QCSHQD, outlining the methodologies and approaches to be employed. Finally in Section \ref{sec:conclusion}, we draw a comprehensive conclusion and our plans.

\section{Background And Motivation}
\label{sec:motivation}

The word quantum in "quantum computing" indicates the quantum mechanics that a system utilize to run computational intensive operations \cite{zhao2020quantum,nielsen2010quantum}. In Physics, quantum is the smallest unit of any physical entity and generally it refers to atomic or subatomic particles e.g., electrons, neutrons and photons. QC leverages the principles of quantum mechanics to process information and perform specific computational tasks much faster than conventional computer systems and allows practitioners and researchers, to probe many possibilities simultaneously.  QC is well suited for tasks like optimization, analysis, simulation, cybersecurity, cryptography, and molecular modelling etc. \cite{zhao2020quantum,mcardle2020quantum,bova2021commercial}.

The key difference between the classical and quantum computer is the qubit \cite{zhao2020quantum,kitaev2002classical}. Differently than classical computer exclusive bits (0,1), qubits can be in a state that is a
superposition of 0 and 1. Superposition allows for the representation of qubits as a linear combination of two basis states |0 and |1. Consequently, any qubit can be represented as \(x|0\rangle + y|1\rangle\), where x and y are complex numbers that adhere to the normalization condition $|x|^2+ |y|^2=1$. This condition ensures that the probability of finding the qubit in either state upon measurement is one. 

The distinct feature of quantum mechanics is the superposition \cite{zhao2020quantum}\cite{dirac1981principles}. Superposition means that a quantum system at a time could be at all possible states rather than one specific state. In quantum computers, a quantum register exists in all possible 0s and 1s superposition states, unlike classical, where the register has only one value at a time. Therefore, two qubits combine as a whole in a superposition of four quantum states, i.e. |00|01|10 and |11, where n qubits can be described as the superposition state of qubits. It is the significant advantage of quantum computers over classical ones where n bits have a fixed single state \cite{zhao2020quantum}.

Designing quantum applications is particularly challenging due to the inherent characteristics of quantum mechanics like superposition and entanglement \cite{horodecki2009quantum}. Therefore, a novel scientific field, quantum software engineering (\textbf{QSE}) \cite{zhao2020quantum,piattini2021toward,piattini2020talavera}, has emerged and focus on fostering the application of traditional software engineering methods to quantum software development. A key point of QSE is integrating classical and quantum systems, which build the so-called hybrid systems \cite{piattini2020talavera,zhao2020quantum}. The requirement for coexistence and collaboration between classical and quantum infrastructures has led to the emergence of hybrid systems. \cite{ramouthar2023hybrid}. In hybridization, specific and complex tasks are assigned to quantum devices, while the less demanding components to classical computing systems, which minimise the risk of unstable and unreliable applications \cite{weder2020integrating,weder2021hybrid}. However, this hybrid concept of quantum and classical computing poses various challenges to application developers, who are required to not only effectively exploit the potentials of quantum hardware and algorithms but also ensure the development of industry-strengths applications \cite{ahmad2023engineering,weder2022quantum,lubinski2022advancing}. It is reasonable to assume that most classical developers and designers might lack the essential skills and expertise to perform low-level plumbing activities in quantum computing infrastructures. They need to go through an intensive learning process to understand pulses, circuits, hamiltonian parameters, and complex algorithms. Approaches focused on workflows, such as those mentioned by \citet{weder2020integrating}, are crucial for effectively orchestrating classical applications with quantum circuits. However, these approaches have limitations in their capacity to systematically reuse modular and parameterized knowledge and implement a structured lifecycle model.
Moreover, they typically do not consider the specific context of applications and often overlook factors of quality and resource constraints. What is required are the methodologies and tools that enable the design, development, deployment, monitoring, and management of applications which can be executed on quantum infrastructure without an in-depth understanding of the underlying principles of quantum mechanics. At the same time, the applications should be dynamically adaptable and reliable, capable of handling (unexpected) changes in the underlying quantum hardware infrastructure. We can address the hybridization challenges by leveraging service-oriented computing/architectures\cite{ahmad2023engineering,ahmad2023reference,papazoglou2008service,PapazoglouTDL07}.
\begin{figure*}[htbp]
\centering
    \includegraphics[width=1.9\columnwidth]{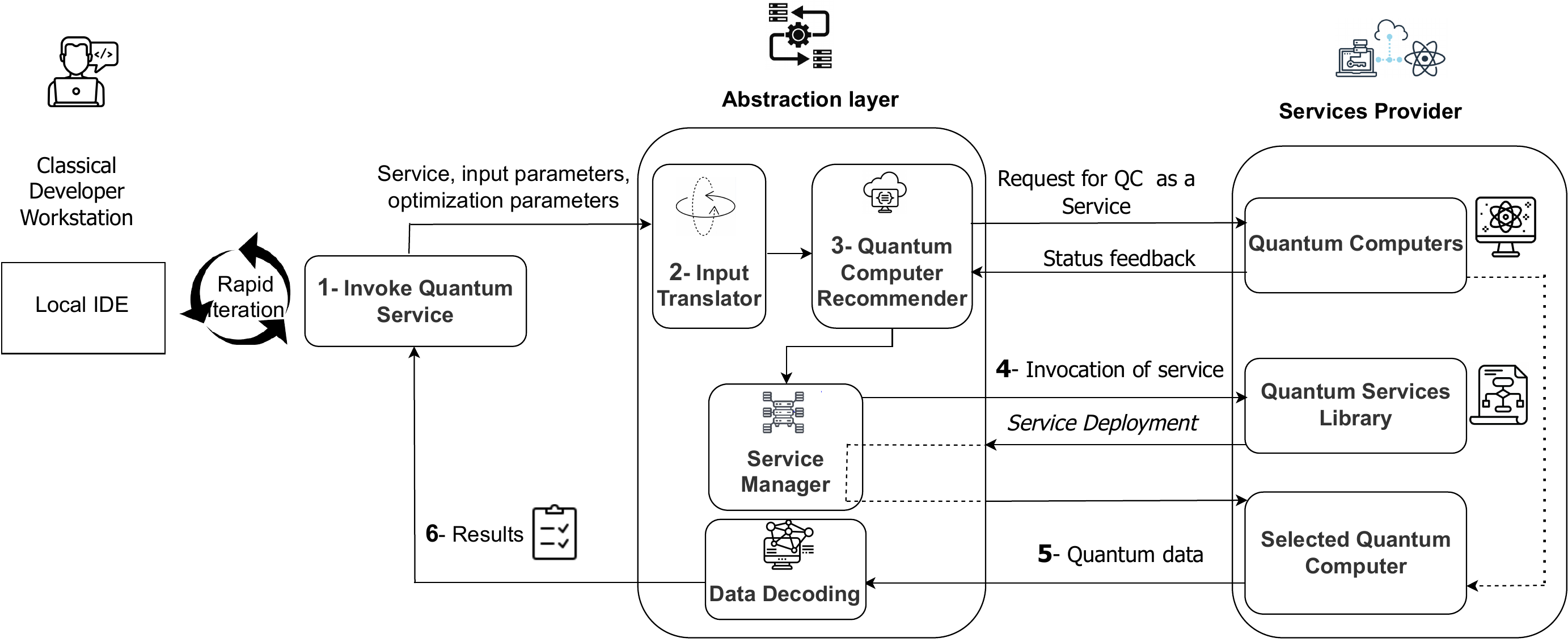}
    \caption{Proposed QCSHQD Framework}
    \label{fig:framework}
\end{figure*}

Service oriented architecture (SOA) is an architectural style in which business and IT systems are designed in terms of services available at an interface and the outcomes of these services. A service is a logical representation of a set of activities that has specified outcomes, is self-contained, it may be composed of other services but consumers of the service need not be aware of any internal structure \cite{iso_soa}.

Following the SOA principles, we introduce a framework aimed at exploiting the potential of SOA within the QC domain. QCSHQD enables users, i.e., classical developers, to locate and employ various QC capabilities, which consist of computing resources and applications provided by service providers. It offers a strategy for developing quantum applications by exploiting a set of reusable functional units, which have well-defined interfaces and are implemented using classical and quantum hybrid computing methodologies.

We now hypothesize a motivating scenario to elaborate the challenges classical developers face to leverage QC, and also to highlight the importance of our scientific endeavours. This scenario also aims to show the impacts and benefits of QCSHQD.

\textbf{Motivating Scenario}: Drug discovery is one of the most significant industries in health. Various pharmaceutical companies are trying to develop life-changing medicines for financial purposes or to improve human life experiences. To this end, they need to simulate molecular interactions and solve optimization problems, which consume a lot of time and energy with classical computers due to the complex nature of chemical processes. While the engineers working in this companies know that by using quantum computers, these simulations can be done with higher accuracy and speed, they will face two significant challenges:
\begin{itemize}
\item Complexity of Quantum Computing: The team lacks expertise in quantum mechanics and finds the quantum computing landscape more complex than classical computing. This gap in knowledge makes it difficult for them to leverage quantum computing for their project.
\item Accessibility of Quantum Resources: Even if they had the necessary knowledge, gaining access to quantum computing resources is another hurdle. Quantum computers are not widely accessible, and interfacing with them requires specific tools, methods, and processes.
\end{itemize} 
We introduce QCSHQD to address such challenges. Our framework offers a solution to make QC resources available as a service to classical developers, eliminating the need for them to have expertise in quantum mechanics. Our framework strives for making QC practical and easily accessible within software development scenarios.

\section{The Vision}
\label{sec:vision}

We introduce QCSHQD, a middleware allowing developers to use QC services in their current classical software development setup. QCSHQD aims to simplify the process of connecting to and using quantum services, so developers can delegate complex computational tasks to QC.

QCSHQD consists of three key components: the Local Integrated Development Environment (IDE) - a dedicated space for classical developers to work; the abstraction layer- responsible for managing quantum services; and the service provider- primarily tasked with the execution of the invoked services.
 Figure \ref{fig:framework} presents QCSHQD workflow, structured in six steps, to streamline access to QC services for users. According to our design, the workflow starts when the developers need the available QC services, which they iteratively invoke using the local IDE interface (1) by defining the input parameters and the optimization parameters of the service (See Figure \ref{fig:framework}). Next based on the specified parameters, the abstraction layers takes these information and translate using input translator (2) into a form that quantum computer can process. Based on the translated parameters, the quantum computer recommender (3) identifies the quantum computer that is available and optimal to execute a given request \cite{valencia2021hybrid}. Once the quantum computer is selected, the service manager requests the required information to deploy the service to the selected quantum computer (4). After the execution of the service, the abstraction layer receives the response (quantum data) from the quantum computer (5) and decode/translate the quantum results back into a classical format (6), as shown in Figure \ref{fig:framework}, which the developers can access using the IDE interface.

For the interaction across the above components, the abstraction layer functions as an intelligent intermediary/middleware and simplifies the complexity of hybridization between classical and quantum computers (See Figure \ref{fig:framework}).

\section{Implementation and Execution Roadmap}
\label{sec:Roadmap}
This section details the tools and methodologies for implementing QCSHQD. We plan to extend the PyDev plugin\footnote{\url{https://www.pydev.org/}} for Eclipse to develop the IDE. Furthermore, for version control and research data management, the project will integrate Git \footnote{\url{https://git-scm.com/}}, which can be effectively used within the Eclipse environment through existing Git plugins.

The abstraction layer design is based on the work of \citet{valencia2021hybrid} and \citet{weder2021hybrid} through OpenAPI\footnote{\url{https://www.openapis.org/}} and TOSCA-based orchestration \cite{lipton2017oasis}. The role of a TOSCA-based orchestration mechanism, \cite{weder2021hybrid,lipton2017oasis} will be automating the process of translating classical input to quantum-compatible formats, selecting appropriate quantum computers, handling the deployment of quantum computing services, interpreting the quantum data and translating them back into classical representation of the results.
The OpenAPI offers a standardized specification for developing and generating APIs to ensure seamless task execution and data integrity. The importance of OpenAPI not only can be defined in users' understanding and exploring services' capabilities, but it will also facilitate services development. Its specification needs to be modified to support quantum services \cite{romero2023enabling}, therefore, we plan to utilize OpenAPI with an extension, considering custom properties tailored specifically for defining quantum applications. These custom properties serve as a means to incorporate supplementary information into the API contract definition, expanding the scope of the specification \cite{romero2023enabling}.

Our primary deployment targets are QC services available on the cloud as well as real-world QC infrastructure, such as VTT quantum computer HELMI connected with CSC LUMI\footnote{\url{https://www.csc.fi/lumi}} for QC provider, which depends on the evolving situation of QC.

Future efforts will be put into defining and evaluating the existing process for developing quantum software based on the QCSHQD. To anticipate this task, we benefit from continuous software engineering practices for team collaboration, short development iterations, and continuous delivery \cite{fitzgerald2017continuous}.

\subsection{Potential Limitations }

QC is still a new domain, and technologies are quickly evolving. Therefore, in this work, we foresee a set of possible scenarios that might impact the proposed road-map. 

\begin{itemize}
    \item The \textbf{integration} of quantum computing services with classical systems may encounter compatibility issues due to the quantum infrastructure readiness.
\item \textbf{Allocation of quantum resources} in a cloud-based service model could be complex, in case of high demand for quantum computing power.

\item \textbf{Scalability: }As the demand for quantum computing resources grows, the framework might face scalability issues.
\item \textbf{Evolution of Quantum Technologies:} The rapidly evolving nature of quantum computing technologies may outpace the development of the proposed framework, necessitating continuous updates and adaptations.

\end{itemize}

\section{Conclusion}
\label{sec:conclusion}

We have introduced QCSHQD, aiming to broaden access to QC resources for classical developers. Through adopting service-oriented computing strategies, QCSHQD streamlines the integration of QC capabilities and reduces the complexity associated with quantum programming. The proposed roadmap for QCSHQD implementation employs modern tools, positioning QCSHQD as a bridge that enables classical developers, regardless of their QC expertise, to exploit the extensive capabilities of QC resources.
Future efforts will focus on implementing QCSHQD, followed by conducting a thorough evaluation of its efficiency and practical applicability. Future efforts will also be put into investigating the impact of our QCSHQD framework on the existing quantum software development process.

\printbibliography

\end{document}